\documentclass[acmsmall,nonacm]{acmart}

\usepackage{graphicx}

\usepackage{longtable}

\usepackage{rotating}
\usepackage{colortbl}
\usepackage{hhline}

\usepackage{lscape}

\usepackage{tikz}
\newcommand{\tikzcircle}[2][red,fill=red]{\tikz[baseline=-0.5ex]\draw[#1,radius=#2] (0,0) circle ;}%

\definecolor{blue}{RGB}{0, 0, 0}

\usepackage{pifont}

\usepackage[font=normalsize]{caption}     
\usepackage{etoolbox}

\AtBeginDocument{%
  \providecommand\BibTeX{{%
    \normalfont B\kern-0.5em{\scshape i\kern-0.25em b}\kern-0.8em\TeX}}}

\setcopyright{acmcopyright}
\copyrightyear{2021}
\acmYear{2021}
\acmDOI{10.1145/1122445.1122456}

\acmConference[Woodstock '18]{Woodstock '18: ACM Symposium on Neural
  Gaze Detection}{June 03--05, 2018}{Woodstock, NY}
\acmBooktitle{Woodstock '18: ACM Symposium on Neural Gaze Detection,
  June 03--05, 2018, Woodstock, NY}
\acmPrice{15.00}
\acmISBN{978-1-4503-XXXX-X/18/06}

\acmJournal{TOMM}
\acmVolume{9}
\acmNumber{9}
\acmArticle{9999}
\acmMonth{9}

\begin{document}


\title[Synthesising Privacy by Design Knowledge Towards Explainable IoT Application Designing in Healthcare]{Synthesising Privacy by Design Knowledge Towards Explainable Internet of Things Application Designing in Healthcare}


\author{Lamya Alkhariji}
\affiliation{%
	\institution{Cardiff University, UK}
}
\email{AlkharijiLa@cardiff.ac.uk}

\author{Nada Alhirabi}
\affiliation{%
	\institution{Cardiff University, UK}
}
\email{alhirabin@cardiff.ac.uk}

\author{Mansour Naser Alraja}
\affiliation{%
	\institution{Dhofar University, Oman}
}
\email{malraja@du.edu.om}

\author{Mahmoud Barhamgi}
\affiliation{%
	\institution{Claude Bernard Lyon 1 University, France}
}
\email{mahmoud.barhamgi@liris.cnrs.fr}

\author{Omer Rana}
\affiliation{%
	\institution{Cardiff University, UK}
}
\email{RanaOF@cardiff.ac.uk}

\author{Charith Perera}
\affiliation{%
	\institution{Cardiff University, UK}
}
\email{pereraC@cardiff.ac.uk}

%

\renewcommand{\shortauthors}{Alkhariji, et al.}

\begin{abstract}
Privacy by Design (PbD) is the most common approach followed by software developers who aim to reduce risks within their application designs, yet it remains commonplace for developers to retain little conceptual understanding of what is meant by privacy. A vision is to develop an intelligent privacy assistant to whom developers can easily ask questions in order to learn how to incorporate different privacy-preserving ideas into their IoT application designs. This paper lays the foundations toward developing such a privacy assistant by synthesising existing PbD knowledge so as to elicit requirements. It is believed that such a privacy assistant should not just prescribe a list of privacy-preserving ideas that developers should incorporate into their design. Instead, it should explain how each prescribed idea helps to protect privacy in a given application design context—this approach is defined as \textit{ ‘Explainable Privacy’}. A total of 74 privacy patterns were analysed and reviewed using ten different PbD schemes to understand how each privacy pattern is built and how each helps to ensure privacy. \textcolor{blue}{ Due to page limitations, we have presented a detailed analysis in \cite{Alkhariji2020}}. In addition, different real-world Internet of Things (IoT) use-cases, including a healthcare application, were used to demonstrate how each privacy pattern could be applied to a given application design. By doing so, several knowledge engineering requirements were identified that need to be considered when developing a privacy assistant. It was also found that, when compared to other IoT application domains, privacy patterns can significantly benefit healthcare applications. In conclusion, this paper identifies the research challenges that must be addressed if one wishes to construct an intelligent privacy assistant that can truly augment software developers’ capabilities at the design phase.
\par 

\end{abstract}



\maketitle

\section{Introduction}
\label{sec:Introduction}

Internet of Things (IoT) \cite{ZMP007} applications aim to collect large volumes of data about us that can be used to derive very sensitive information about our way of life and behaviours. Specifically, healthcare IoT applications \cite{Chen2007} may collect and derive very sensitive information about end-users (e.g. patients). The immediate implications of this are that IoT systems are inherently a threat to personal data privacy. Privacy violations could be a serious charge to be held against IoT systems. It has been found out that a significant number of people change their mind about using a system/application once they become aware of the data collected about them \cite{Hong2017}. On this account, it is important to develop IoT applications in a privacy-aware manner. One major barrier is the lack of privacy knowledge among developers \cite{PrivacyAssessment,Perera2017} \footnotetext{In this paper, we use the term \textit{software developer} to broadly represent all types roles in the software development life cycle such as front end developers, back end developers, software designers, architects, etc. We acknowledge that in larger organisations, each of these roles has very specific job responsibilities where each developer focus on a specific aspect. However, in small organisations, software developers may involve all tasks ranging from requirement engineering, software design, coding, and testing. We decided to use the term \textit{software developer} as  differences between roles does not make any difference to the discussion of this paper.}. To make this issue worse, IoT applications are complex systems \cite{Sivaram2009} where both software and hardware components(e.g., sensors and actuators) need to work together across many different types of nodes (e.g., microcontrollers, system-on-a-chips, mobile phones, miniaturised single-board computers, cloud platforms) with different capabilities under different conditions.

It is well known that software developers invest most of their time and effort on functionalities. Non-functional requirements such as privacy and security are typically considered as an afterthought \cite{Wang2018, Sakai2013, Wu2016}. There are many reasons for such behaviour, ranging from lack of knowledge (e.g., privacy knowledge) through to cost and time pressures. It is believed that the best way to address this problem (i.e., lack of focus on non-functional requirements such as privacy) is to augment software developers with tools and techniques that help them to achieve their job both efficiently and effectively. Such an approach should be less dependent on the knowledge of software developers, and furthermore, should allow them to achieve tasks easier and faster, meaning that they do not need to compromise either cost and time that they would otherwise spend on developing functional requirements. To this end, the notion is proposed herein that a privacy assistant is a good way to address this issue.

In order to understand the behavioural characteristics of a privacy assistant, ten expert and novice software developers were interviewed \cite{Perera2017}. It was found that software developers do not like to be told what to do. Notably, the more senior (or experienced) they become, the more reluctant they are to blindly follow prescriptive advice provided by an intelligent tool. Instead commenting that a tool designed to help them understand privacy, yet which allowed them to retain critical judgement on the implementation of privacy preserving ideas in their work. What they are more interested in knowing is ‘why and how a particular privacy-preserving idea helps to protect privacy and reduce privacy risk in a given application design’. In this paper, this concept is defined \textit{ ‘Explainable Privacy’}. The exact definition of this phrase will be considered in detail later.

\subsection{Contribution}
It was identified that, in order to develop an intelligent and explainable privacy assistant, the extant PbD knowledge must first be synthesised. In other words, the knowledge must be collated and brought together in order to offer an all-encompassing model which can ably describe the systems of interest. This is the main focus of this paper, the contributions of which are as follows:
\begin{itemize}
	\item \textcolor{blue}{
The extant literature was surveyed, with ten well-known PbD schemes selected and 74 well-known privacy patterns extracted. Each privacy pattern was examined against these PbD schemes \footnotetext{Different PbD schemes use different terminology such as guidelines, principles, strategies goals, rights, to refer to the elements within schemes.} . Thus, it was possible to explain how each privacy pattern was developed by combining different PbD schemes in a way which is meaningful to software developers. Due to page limitations, only two of those analyses are presented. The full analysis can be found in \cite{Alkhariji2020} .}
\item \textcolor{blue}{
Demonstration of how privacy patterns could be mapped to three real-world given IoT application designs is provided.Due to page limitations the third use case is in the appendix. This mapping process allowed the extraction of several knowledge modelling requirements. It was found that privacy patterns are useful in non-healthcare application scenarios, but less so compared to the healthcare applications. }
\item \textcolor{blue}{
Several research challenges were identified—these need to be addressed while building an intelligent privacy assistant that can augment software developers’ capabilities at the design phase. A discussion of how the knowledge synthesis contained herein can address these research challenges is provided. }
\end{itemize}

\subsection{Review of some survey papers }

\textcolor{blue}{
Many researchers have been working on improving privacy in IoT applications. For example, Li et al. \cite{li2017hierarchical} have proposed deploying a two-layer and three-layer game model to preserve user data in IoT applications. This is a game that models not only the user and attacker interactions, but also the interactions between the user, attacker, and service provider. This game offers a novel hierarchal framework that helps to model interactions between the three layers, demonstrates the impact of the service provider on the user-attacker interaction, and guides the service provider whilst making privacy policies. However, this framework can only be implemented during the run phase. Where this paper emphasises the PbD concept and privacy preserving practices are implemented in the design phase. 
}
\par
\textcolor{blue}{
Hassan et al. \cite{hassan2019privacy} reported briefly on common IoT systems, including blockchain-based IoT systems, which are used for financial functionalities. Subsequently, they discuss five privacy preserving strategies, but in doing so they blended security with privacy issues. In addition, their recommendations are limited to application in blockchain IoT systems. In this study, more varied privacy preserving schemes are collated, showing a total of 161 scheme elements.
}
\par
\textcolor{blue}{
Abdul-Ghani and Konstantas \cite{abdul2019comprehensive} defined guidelines for privacy and security in IoT applications using edge nodes and communication layers. They also stated who these guidelines are intended for i.e. Manufacturer, Developer, Customer, and Provider. They chose a group of guidelines for both security and privacy and showed who and how to implement them. Whereas this work combines PbD schemes to give structure to the available privacy preserving practices, whilst simultaneously showing how they can be implemented through IoT use cases, to exemplify the proposed tooling. In summary, this work provides more in depth coverage of PbD philosophy, highlights the hierarchical nature of the extent of privacy levels offered within a scheme, and explores the links between them.  }

\subsection{Paper structure}

This paper is organised as follows: Section \ref{sec:PrivacybyDesignKnowledge} introduces the types of PbD Schemes and their differences, before summarising the analysis techniques used to assess each privacy pattern. Section \ref{sec:PrivacybyDesignforInternetofThings} depicts how the aforementioned privacy patterns can be mapped into reality through the provision of several IoT use cases. Subsequently, section \ref{sec:DiscussionandLessonsLearned}, discusses several knowledge modelling requirements with help from the findings in sections \ref{sec:PrivacybyDesignKnowledge} and \ref{sec:PrivacybyDesignforInternetofThings}. Section \ref{sec:ResearchChallengesandFutureDirections} identifies future research direction and challenges. Concluding remarks offered in section \ref{sec:Conclusion}.

\section{Privacy by Design Knowledge}
\label{sec:PrivacybyDesignKnowledge}

Privacy is a vague and fuzzy concept without a rigid definition. It is widely misunderstood that privacy can only be protected by securing or encrypting the data itself \cite{gdpr_2020}, as in Microsoft’s STRIDE model \cite{anwar2020modeling} \cite{dewitte2019comparison}. In fact, the design of a system could play a large role in preserving data privacy, which leads to the concept of Privacy by Design (PbD). It is an approach to systems development initially proposed by Cavoukian \cite{ANN} in 1995, after her long experience in the Office of the Information and Privacy Commissioner. It is a proactive approach that considers data privacy from the early stages of the development phase \cite{wong2019bringing}.

Over time, different parties have proposed different privacy by design (PbD) schemes (e.g., privacy principles, strategies, guidelines, and patterns) to guide software designers and developers in the creation of privacy aware systems. Due to the variations in originating contexts (i.e., different people with different aims in mind and different backgrounds), each of these schemes is designed to stand alone, without proper connection to the others. Such a disconnected view makes it difficult, confusing, and frustrating for anyone to understand and make use of different PbD Schemes (or to propose new schemes to mitigate existing gaps). This section first explores the extant literature covering the foundations of PbD, seeking to identify, analyse, extract meaning, and ultimately synthesise independently built privacy patterns into a comparative structure capable of assessing the validity and applicability of these existing proposals.

\subsection{Privacy Knowledge Hierarchy}
\label{sec:2}

Cavoukian’s initial PbD framework proposed seven foundational principles. Although the principles are fairly easy to understand, it requires significant effort to transform these principles into implementable components within the software development process. Therefore, software developers often find them high-level and less useful in their day-to-day work. As a consequence, relatively few PbD systems have been proposed either by industry or academia. Table \ref{Tbl:ExistingPrivacybyDesignSchemes} summarises the ten well-known PbD schemes to be considered. As mentioned earlier, it is important to note that literature on PbD schemes uses a number of terms such as \textit{ principle, strategies} and \textit{patterns}, to refer to the elements within each scheme. The table is ordered by Type and Year.

\begin{table}[h!]
	\centering
	\caption{Existing Privacy by Design Schemes}
	\label{Tbl:ExistingPrivacybyDesignSchemes}       

	\begin{tabular}{llll}
		\hline\noalign{\smallskip}
		Citation & Year &Type & Number  \\
		\noalign{\smallskip}\hline\noalign{\smallskip}
		
	Fair Information Practice Principles (FIPPs) \cite{Cate2006} & 2006 & Principles    & 5                                                               \\
	Cavoukian \cite{Cavoukian}                              & 2009 & Principles    & 7                                                                \\ 
		
	ISO 29100 Privacy framework \cite{ISO2015}         & 2011 & Principles    & 11                                                                 \\
		
	Cavoukian and Jonas \cite{Cavoukian2012}           & 2012 & Principles    & 7    \\

	Wright and Raab \cite{Wright2014}                & 2014 & Principles    & 9                                                                \\
	Fisk et al. \cite{Fisk2015}                           & 2015 & Principles    & 3                                                                 \\
	
	Hoepman \cite{Hoepmanraey}       & 2018 & Strategies    & 8                                                                 \\
	
	Rost and Bock \cite{Rost2011}                       & 2011 & Goals         & 6                                                                 \\
	Economic Cooperation and Development (OECD) \cite{Oleary1995}        &1995        & Guidelines    & 8                                                                 \\ 
		 
	Perera et al.  \cite{Perera2017}                       & 2017 &Guidelines    & 30                     \\ 
	
		\noalign{\smallskip}\hline
	\end{tabular}
\end{table}

\begin{figure}[t!]
	\centering
	\includegraphics[scale=0.19]{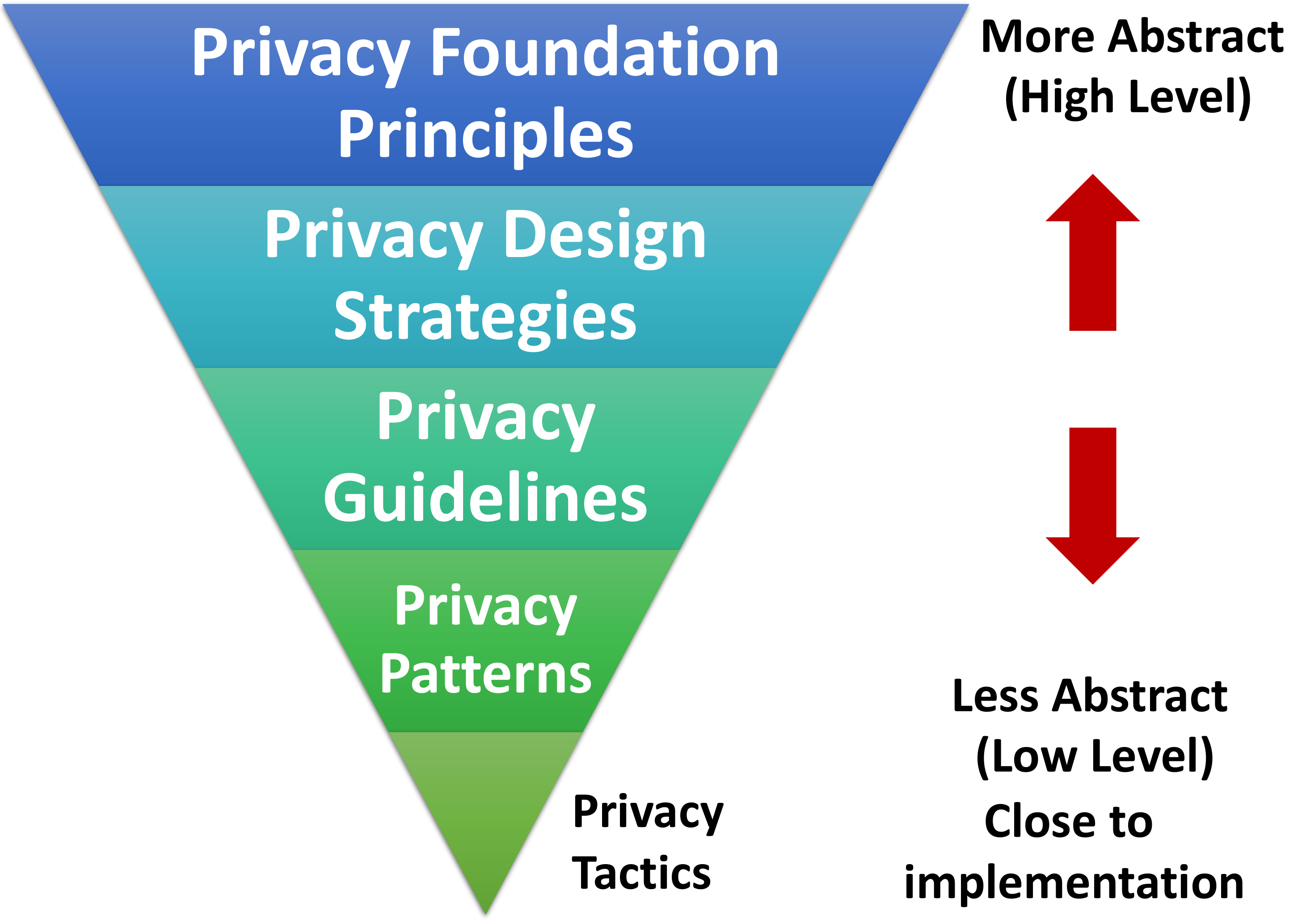}
	\caption{From high level principles to low-level tactics:  \textcolor{black}{Rubinstein and Good \cite{Rubinstein2013}  argue that making a specification or requirement for  software design is to make it concrete, specific, and preferably associated with a metric. The layered approach aims to achieve this in a systematic way.}}
	\label{Figure:BigPicture}	
\end{figure}

Despite the absence of agreement within the extant literature, approximate relationships can be drawn between these terms (see Figure \ref{Figure:BigPicture}). Principles can be considered to represent more abstract, higher level, ideas. In contrast, tactics are low-level, concrete instructions for implementing solutions in a specific context. Strategies, guidelines and patterns sit between them. This does not mean one type is better or worse than the others, since each layer has its own strengths and weaknesses. However, it is acknowledged that the boundaries between these layers are soft, to the extent that some principles may be interpreted as strategies and vice-versa.

\begin{description}
	\item[Principle:] A principle is a concept or value that is a guide for behaviour or evaluation. Typically, they are very abstract and provide an overall direction to follow. Ten Privacy Principles of Personal Information Protection and Electronic Documents Act (PIPEDA) \cite{MinisterofJustice} and the document Seven Foundation Privacy by Design principles, written by the Information \& Privacy Commissioner of Canada \cite{Cavoukian} are key examples.

	\item[Strategies:]  In contrast to principles, strategies are focused on achieving a specific outcome. A design strategy describes a fundamental approach to achieve a certain design goal. Therefore, strategies are more specific in terms of what they aim to achieve. Hoepman’s \cite{Hoepmanraey} seven privacy design strategies is an example.
	
	\item[Guidelines:] The guidelines adopted in this paper break down strategies into a lower-level, concrete set of instructions that a software engineer can follow.
	
	\item[Patterns:] Design patterns are useful for making  decisions about the organisation of a software system. A design pattern \textit{``provides a scheme for refining the subsystems or components of a software system, or the relationships between them. It describes a commonly recurring structure of communicating components that solves a general design problem within a particular context.''} \cite{Buschmann1996}. Patterns solve a specific problem but are neutral or have weaknesses with respect to other qualities. In contrast, there are also \textit{`anti-patterns'}. In software engineering, an anti-pattern is a design that may be commonly used but is ineffective or counterproductive in practice \cite{Budgen2003}.

\end{description}

It is important to note that the top three layers (principles, strategies, and guidelines) primarily adopt a top-down approach. They suggest good practice and have been historically or logically proven to reduce privacy risks. Typically, they are a blanket solution that aims to eliminate multiple privacy issues simultaneously (without addressing them individually). In contrast, patterns focus on solving specific problems. This is more of a bottom-up approach to identifying solutions to specific privacy problems.

These layers are perhaps best explained through the use of an example which can simultaneously highlight the weakness of the boundary layers. \textbf{[Principle]} \textit{``Proactive not Reactive; Preventative not Remedial ''}  is one of the principles proposed by the Canadian Information \& Privacy Commissioner \cite{Cavoukian}. The official explanation of which is that the \textit{``Privacy by Design (PbD) approach is characterised by proactive rather than reactive measures. It anticipates and prevents privacy-invasive events before they happen. PbD does not wait for privacy risks to materialise, nor does it offer remedies for resolving privacy infractions once they have occurred—it aims to prevent them from occurring. In short, Privacy by Design comes before-the-fact, not after ''}.

Application of this principle thus allows the development of a \textit{`Minimise'} \textbf{[Strategy]}, which Hoepman describes as \textit{``limiting usage as much as possible by excluding, selecting, stripping, or destroying any storage, collection, retention or operation on personal data, within the constraints of the agreed-upon purposes ''} \cite{Hoepmanraey}. Hoepman’s minimise strategy can be identified as a way to follow the \textit{`proactive principle '} (i.e. minimise the amount of data collected as a proactive measure to avoid or reduce potential privacy violations).

It is possible to further break down the minimise strategy into guidelines. \textbf{[Guideline]} One minimisation guideline is to \textit{``minimise raw data intake ''}, which can be further defined as \textit{`` whenever possible, IoT applications should reduce the amount of raw data intake. Raw data could lead to secondary usage and privacy violation. Therefore, IoT applications should consider converting (or transforming) raw data into secondary context data ''}  \cite{PrivacyAssessment}.

A privacy pattern can be identified as a low-level design that aims to solve a specific privacy challenge. The relationship between guidelines and patterns may be quite weak as, in most instances, patterns can stand by themselves as problem solving techniques. However, privacy patterns can still be identified as low-level designs that help to implement guidelines. Continuing the example, a \textbf{[pattern]} could be extracted called an \textit{`Online Activity Detector '}. This pattern extracts orientation (e.g. sitting, standing, walking) by processing accelerometer data and only stores the results (i.e. secondary context) and deletes the raw accelerometer data.

\subsection{Synthesising Methodology}
\label{subsec:SynthesisingMethodology}

There are significant similarities and differences between PbD Schemes. It was decided to examine and analyse them further with the intention of uncovering the relationships between different PbD schemes. After careful consideration, patterns were selected for further analysis, since \textit{Patterns} are the closest to implementation (i.e., least abstract). An exhaustive literature search, allowed the compilation of a list of 74 patterns \cite{Perera2020}. These patterns were primarily adopted from two main sources \cite{privacypatterns.eu,privacypatterns.org}, with secondary assessment conducted to remove duplicates before undertaking simultaneous cross-examination of each listed pattern against alternative PbD schemes and synthesis by two independent researchers. These results were then collated through a process of rigorous debate.

Due to page limitations, only two of these analyses are presented herein, with the full analysis available in \cite{Alkhariji2020}. As representative samples, one principle-based PbD scheme (ISO 29100 Privacy framework \cite{ISO2015}) and one guideline-based PbD scheme (Perera et al. \cite{Perera2017}) are presented. Table \ref{Tbl:ISOPrivacyPrinciples} compares privacy patterns with the privacy principles presented in the ISO 29100 Privacy Framework, while Table \ref{Tbl:HoepmanPrivacyStrategies} contrasts privacy patterns with the privacy strategies proposed by Hoepman \cite{Hoepmanraey}. In these tables, green circles ( \tikzcircle[black!60!green, fill=black!60!green]{1.5pt}  ) show that a given pattern is fully inspired by the given PbD element (i.e. principles/guidelines), whereas orange circles ( \tikzcircle[orange, fill=orange]{1.5pt} ) indicate that the PbD element partially inspired a given pattern. A more detailed example of this analysis process is provided later in this paper.

\subsection{Example Privacy Pattern Analysis: Location Granularity}
\label{subsec:ExamplePrivacyPatternAnalysis}

In order to fully describe the examination process, it is necessary to first detail the Location Granularity pattern using the definitions made in \cite{privacypatterns.eu,privacypatterns.org}.

{ \begin{description}
		\item[\textbf{Name}] \textit{Location Granularity}
		
		\item[\textbf{Summary}] \textit{Support minimization of data collection and distribution. Important when a service is collecting location data from or about a user, or transmitting location data about a user to a third-party.}
		
		\item[\textbf{Problem}] 
		
		\textit{Many location-based services collect current or ongoing location information from a user in order to provide some contextual services (nearest coffee shop; local weather; etc.). Collecting more information than is necessary can harm the user's privacy and increase the risk for the service (in the case of a security breach, for example), but location data may still need to be collected to provide the service. Similarly, users may want the advantages of sharing their location from your service to friends or to some other service, but sharing very precise information provides a much greater risk to users (of re-identification, stalking, physical intrusion, etc.).}

		\textit{Accepting or transmitting location data at different levels of granularity generally requires a location hierarchy or geographic ontology agreed-upon by both services and a more complex data storage model than simple digital coordinates.}

		\textit{Truncating latitude and longitude coordinates to a certain number of decimal places may decrease precision, but is generally not considered a good fuzzing algorithm. (For example, if a user is moving in a straight line and regularly updating their location, truncated location information will occasionally reveal precise location when the user crosses a lat/lon boundary.) Similarly, using "town" rather than lat/lon may occasionally reveal more precise data than expected when the user crosses a border between two towns.}
		
		\item[\textbf{Solution}] \textit{Since much geographic data inherently has different levels of precision like street, city, county, state, country -- there may be natural divisions in the precision of location data. By collecting or distributing only the necessary level of granularity, a service may be able to maintain the same functionality without requesting or distributing potentially sensitive data. A local weather site can access only the user's zip code to provide relevant weather information without ever accessing precise (and therefore sensitive) location information.}
		
\end{description}}

Once a specific privacy pattern was understood, it could be compared against each element of all ten different PbD schemes shortlisted for this study. A few selected samples are now presented to demonstrate this decision-making process. It is important to note that, on several occasions, judgement calls were made through rigorous debate between authors, which was centred on the descriptions of each PbD scheme. \textcolor{blue} { A green circle ( \tikzcircle[black!60!green, fill=black!60!green]{1.5pt} ) was chosen where the descriptions were found to seek to the same aim, whilst an orange circle ( \tikzcircle[orange, fill=orange]{1.5pt}  ) was used if the descriptions were less precisely relevant.} However, some decisions remain open for debate and are perhaps controversial depending on how each pattern and PbD element are interpreted. The nature and extent of this controversy are considered in more detail in the discussion.

{\tiny
	\centering
\begin{longtable}{l lllllllllll } 
		\caption{Privacy Patterns Examined Against Privacy Principles \cite{ISO2015}}	
		\label{Tbl:ISOPrivacyPrinciples} \newline
		\setlength\tabcolsep{1pt}
		\\
		Privacy Pattern   List                                                      &
		\begin{sideways}  1.      Consent and choice \end{sideways} & 
		\begin{sideways} \begin{tabular}[c]{@{}l@{}} 2.      Purpose legitimacy \\ and specification \end{tabular} \end{sideways} & 
		\begin{sideways} 3.      Collection limitation  \end{sideways} & 
		\begin{sideways} 4.      Data minimization  \end{sideways} & 
		\begin{sideways} \begin{tabular}[c]{@{}l@{}} 5.      
				Use,retention and \\ disclosure limitation \end{tabular} \end{sideways}  & 
		\begin{sideways} 6.      Accuracy and quality \end{sideways} 
		& 
		\begin{sideways} \begin{tabular}[c]{@{}l@{}}  7.      Openness, transparency \\ and notice \end{tabular} \end{sideways} 
		& 
		\begin{sideways}
			\begin{tabular}[c]{@{}l@{}} 
				8.      Individual participation \\ and access
		\end{tabular} \end{sideways} 
		& 
		\begin{sideways} \begin{tabular}[c]{@{}l@{}}  09. Accountability \end{tabular} \end{sideways} 
		& 
		\begin{sideways} \begin{tabular}[c]{@{}l@{}}  10. Information security \end{tabular} \end{sideways} 
		& 
		\begin{sideways}  \begin{tabular}[c]{@{}l@{}}  	11. Privacy compliance  \end{tabular} \end{sideways}

		\smallskip	
		\\ 
		\rowcolor[rgb]{.951,.951,.951} 1.       Protection against Tracking                                               &                               &                      &  \tikzcircle[orange, fill=orange]{1.5pt}                             &      \tikzcircle[black!60!green, fill=black!60!green]{1.5pt}                  &       \tikzcircle[black!60!green, fill=black!60!green]{1.5pt}                      &                  &                                  &    &&&                     \\
\end{longtable}
}

\begin{description}
	\item[\textbf{Consent and choice (Principle NO:1 in  ISO 29100  Framework \cite{ISO2015})}] \textit{{Location Granularity}} pattern says that systems should provide the opportunity for the users to decide on which granularity level (e.g., As precise as possible, Postal code, neighbourhood, town, region, state, country) they would like the systems to process their data. Therefore, \textit{Location Granularity} is considered to only partially follow the  \textit{Consent and choice} principle, as it does not explicitly handle \textit{Consent and choice}. 
\end{description}

\begin{description}
	\item[\textbf{Purpose Legitimacy (Principle NO:2 in  ISO 29100  Framework \cite{ISO2015})}] \textit{Location Granularity} pattern says that the system should let the users decide the granularity they prefer the system to use. This means that implicitly, the system is receiving the consent to use location data. Therefore, \textit{Location Granularity} is taken to partially adhere to the \textit{Purpose Legitimacy and specification} principle. 
\end{description}

\begin{description}
	\item[\textbf{Minimise Strategy (Guideline NO:1 in  Hoepman \cite{Hoepmanraey})}] It is clear that \textit{Location Granularity} pattern follows this strategy and helps to reduce data quantity.
\end{description}

\begin{description}
	\item[\textbf{Hide Strategy (Guideline NO:2 in  Hoepman \cite{Hoepmanraey})}] 
	The Hide strategy is defined as \textit{Protect personal data, or make it unlinkable or unobservable. Make sure it does not become public or known.} It is believed that the \textit{Location Granularity} pattern partially follows the \textit{Hide} strategy because it helps to reduce linkability. For example, a system that collects exact geo-coordinates can be easily converted to a house address, which can be then linked to an individual. Therefore, a system should only collect less granular data—such as neighbourhood, town, region, state, or country—to reduce linkability.
\end{description}

\section{Privacy by Design for Internet of Things}
\label{sec:PrivacybyDesignforInternetofThings}

This section depicts how privacy patterns reduce privacy violations using real-world IoT case studies (see appendix for third case). Once the privacy patterns have been mapped into the IoT application design, the previous analysis (i.e. synthesised PbD knowledge) can be used to explain how each pattern helps to protect privacy in a given design. Such explanations will help to construct an intelligent Privacy Assistant that can explain its own recommendations.

The three use cases selected cover (1) Real-Time Vehicle Tracking (2) Diabetes
Treatment Monitoring and (3) Performance Evaluation in a Leisure Park. These cases highlight the extreme breadth of complexity and yet shared functionality found in what seem on first inspect to be disparate scenarios having little in common, yet are all ably tackled by IoT applications. During the demonstration, it was found that healthcare applications (Diabetes Treatment Monitoring) are the use case scenario that most greatly benefited from privacy patterns. It was found that \textit{41 (out of 74)} privacy patterns were applicable for Real-Time Vehicle Tracking, whereas \textit{56} patterns were applicable to Diabetes Treatment Monitoring. \textit{Twenty} privacy patterns were applicable for Performance Evaluation in a Leisure Park. It is acknowledged that such applicability of privacy patterns may be subjective due to various factors including, but not limited to, user requirements, the experience of the software developer, and architectural choices. However, it can be argued that healthcare applications benefited from privacy patterns more than the other domains due to the sensitivity of the data collected and processed.


{\tiny
	\centering
\begin{longtable}{p{6.5cm} l llllllll } 
		\caption{Privacy Patterns Examined Against Privacy Strategies  \cite{Hoepmanraey}}	\\	
		\label{Tbl:HoepmanPrivacyStrategies} \newline
		
		Privacy Pattern   List                                                      &
		\begin{sideways}  1.      Minimize \end{sideways} & 
		\begin{sideways} \begin{tabular}[c]{@{}l@{}} 2.      Hide  \end{tabular} \end{sideways} & 
		\begin{sideways} 3.      Separate  \end{sideways} & 
		\begin{sideways} 4.     Aggregate  \end{sideways} & 
		\begin{sideways} \begin{tabular}[c]{@{}l@{}} 5.      
				Inform \end{tabular} \end{sideways}  & 
		\begin{sideways} 6.      Control \end{sideways} 
		& 
		\begin{sideways} \begin{tabular}[c]{@{}l@{}}  7.      Enforce \end{tabular} \end{sideways} 
		& 
		\begin{sideways}
			\begin{tabular}[c]{@{}l@{}} 
				8.    Demonstrate
		\end{tabular} \end{sideways}  	
		\smallskip
		\\ 			
		\rowcolor[rgb]{.951,.951,.951} 1.       Protection against Tracking                                               & \tikzcircle[black!60!green, fill=black!60!green]{1.5pt}                              &                      &                             &  \tikzcircle[orange, fill=orange]{1.5pt}                &                             &    \tikzcircle[black!60!green, fill=black!60!green]{1.5pt}              &            &                                              \\
		2.       Location Granularity                                                      & \tikzcircle[black!60!green, fill=black!60!green]{1.5pt}                               &  \tikzcircle[orange, fill=orange]{1.5pt}                &                              &                        &                        &   \tikzcircle[black!60!green, fill=black!60!green]{1.5pt}               &                                  &                        \\
\end{longtable}
}

\subsection{Use Case 1: Real-Time Vehicle Tracking}

\begin{figure}[!h]
	\centering
	\includegraphics[scale=0.4]{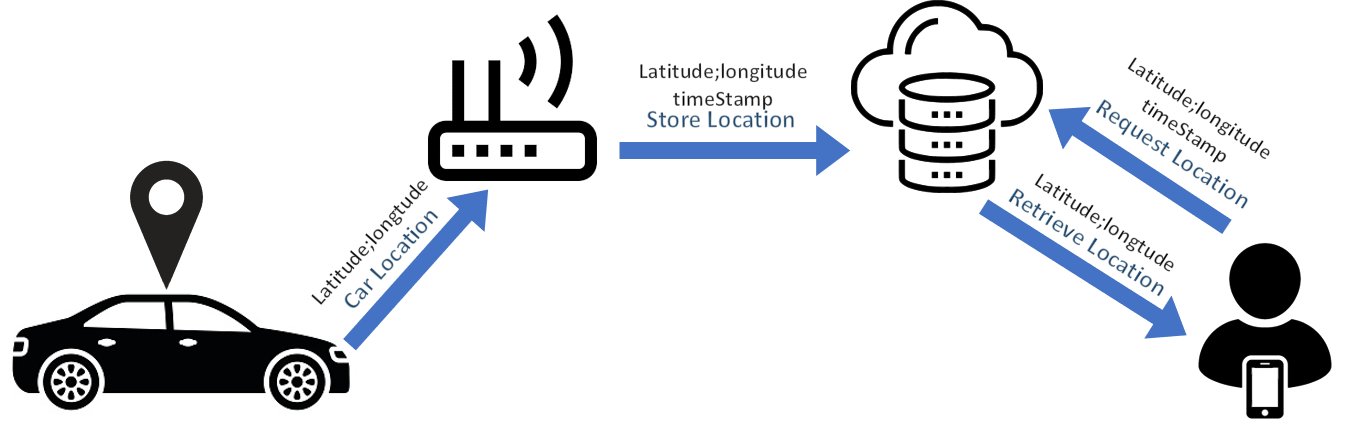}
	\caption{Use case scenario: \textit{CarFinder}, Real-time Vehicle Tracking System}
	\label{Fig:CarFinderScenario}
\end{figure}

\textit{CarFinder}  \cite{IoTCarFinder} is a simple IoT application. It aims to help the driver to find where they parked their car. It is considered as a real time tracking solution as it stores and retrieves the parked car location from a centralised system. As Figure \ref{Fig:CarFinderScenario} illustrates, location data will be sent by a GPS device embedded in the car to be stored in the cloud via a gateway. When the user wants to retrieve this information, they request the location data through a mobile application. At this point, the application will retrieve the car’s location and show it to the user. \textcolor{blue}{ Despite location tracking being commonplace in contemporary society, many developers remain oblivious to the potential breaches in user privacy that may occur during what they blindly consider to be routine data collection for service improvement. It is for this reason that this example was chosen.} Table \ref{Tbl:CarFinderPatterns} explains how each privacy pattern helps to reduce privacy risk in the \textit{CarFinder} IoT application.


{\tiny
	\begin{longtable} {p{4cm} | p{7.3cm}}
		\caption{Application of Privacy Pattern to \textit{CarFinder} IoT Application}\\
			\hline 

		\label{Tbl:CarFinderPatterns} 
		Privacy Pattern   & Example 
		\\ \hline

		%
		
		3. Minimal Information Asymmetry                                                       & The information collected by the service provider should be as minimum as possible. In addition, data collected, and privacy policies should be clearly known to user.                                                                                                                                                                                        \\ 
		
		4. Informed Secure Passwords                                                            & The user should be assisted to maintain a strong password, while singing up, to protect themselves..                                                                                                                                          \\ 
		
		5. Awareness Feed.                                                                      & The user should be informed, via the application, before the application processes the collection of the car locations.                                                                                                                         \\ 
		
		%
		
		8. Use of dummies                                                                       & The system should vague the user’s data by adding fake locations in the database.          \\ 
		
		9. Who's Listening                                                                        &          User should be able to know who have the access to view their data.                                                                                                                                                                                                                \\ 
		
		10. Privacy Policy Display                                                                 & The system should allow users to display the privacy policy it follows.                            \\ 
		
		11. Layered Policy Design                                                                  & The privacy policy should be summarized in a nested way, so users can understand how their data is managed and for which purposes.    \\ 
		
		%
		
		14. Asynchronous notice                                                                  & Users should be provided a simple reminder that they have consented to the sensor to send car location and they could configure the settings if they do not want to see this reminder again.       \\ 
		
		15. Abridged Terms and Condition                                                         & The privacy policy should be summarized in a concise way, so users can understand how their data is managed and for which purposes.      \\ 
		
		16. Policy Matching Display                                                                & The system should retrieve the user’s privacy preferences, via some controller, and compare them to its own policies, then, the contradictions should be highlighted to the user.    \\ 
		
		
		18. Outsourcing [with consent]                                                             & The controller should obtain an additional law consent from users before processing their cars’ locations to third parties.   \\ 
		
		19. Ambient Notice                                                                         & The system should notify users every time the sensor collects the cars’ locations.             \\ 
		
		20. Discouraging Privacy Policy                                                            & In the user interface, a tool tip should be designed, so the user can have a quick glance to the related policy in this input/ service in the interface.           \\ 
		
		21. Privacy Labels                                                                         & The application should use standardized privacy labels to help users understand the privacy policy with less effort.  \\ 
		
		22. Data Breach Notification Pattern                                                       & The application should react to data breach quickly by notifying the user with the Breach details.    \\ 
		
		
		24. Onion Routing                                                                          & Car’s location should be encrypted in many layers, i.e. An encryption layer for the data sent from the car to be decrypted in the gateway, another layer of encryption to be decrypted in the cloud and so on. \\ 
		
		%
		
		27. Personal Data Store                                                                    & The location of the car and all other stored data related to the user should be controlled by the user himself. It could be done by combining a central server and secure personal tokens. \\ 
		
		28. Trust Evaluation of Services Slides                                                    & The application should implement a trust evaluation function to establish a reliable trust between users and controllers.                                                                                                 \\ 
		
		
		30. Privacy icons                                                                          & The application should provide agreed icons beside texts to make privacy policy understood in an easy visual way. \\ 
		
		
		32. Sign an Agreement             & The controller should provide the user with a contractual agreement to make the controller more obligated to their word.  \\ 
		
		%
		
		35. Enable/Disable Function                                                                & Users should be able to enable and disable any functions at any time, i.e. disable the feature of collecting the car location.              \\ 
		
		36. Privacy Color Coding                                                                   & The privacy policies followed by the system should be displayed in a standardized color visual cues, so it helps the user to understand the policy easily.   \\ 
		
		37. Appropriate Privacy Icons                                                              & The application should provide the appropriate icons for each privacy policy, so it helps the user to understand the policy easily.   \\ 
		
		38. User Data Confinement Pattern                                                          & Instead of storing car location to in the system’s cloud, it should be stored locally in the user’s device.  \\ 
		
		39. Icons for Privacy Policies                                                             & In the privacy policy document, icons should be provided for the policies, so it helps the user to understand the policy easily. \\ 
		
		40. Obtaining Explicit Consent                                                             & The system should prompt a clear unambiguous consent from the user before it starts processing his data.               \\ 
		
		41. Privacy Mirrors                                                                        & Provide access to all data stored about the user, he should be able to browse all stored data about himself. \\ 
		
		42. Appropriate Privacy Feedback                                                           & Notify user whenever new data is collected by sending notifications each time, so he would be fulling understanding the privacy risks involved.   \\ 
		
		%
		
		45. Platform for Privacy Preferences                                                       & Applications are encouraged to use similar privacy preferences platforms, so policy distinctions will be highlighted.      \\ 
		
		%
		
		48. Privacy Dashboard                                                                      & In the application, provide easy and full access to the user where he can review his collected data.                      \\ 
		
		
		50. Obligation Management                                                                  &          Car finder system should use an obligation management system, which manipulates data over time, ensuring data deletion and notifications to the user.   \\ 
		
		51. Informed Credential Selection                                                          &   In the application’s interface, a summary of collected data and possible choices should be presented to the user.    \\ 
		
		
		53. Negotiation of Privacy Policy                                                          & The application should ask user if he wants his parking location to be stored by default all the time unless he chooses “No”.    \\ 
		
		54. Reasonable Level of Control                                                            &     The system should allow the user to choose what data he agrees to collect.    \\ 
		
		%
		%
		
		58. Lawful Consent                                                                         &        In the terms and conditions sheet, all data to be collected from the user should be stated clearly to be accepted, giving the user the option of trading off the services once he refuses a certain permission.             \\ 
		
		59. Privacy Aware Wording                                                                  & In privacy policies document, used vocabulary should be easy to understand by any user, considering applying other patterns as well, such as, layered policy design.   \\ 
		
		60. Sticky Policies                                                                        &   User should be able to manage his collected data even in the case that they are shared with a third party via an obligation management system.  \\ 
		
		61. Personal Data Table                                                                    &   The system should provide a platform where the user would be able to keep track of his data along with the process that it goes throw.    \\ 
		
		62. Informed Consent                                             & Display the consent form for the user in a user-friendly way such as, Just-In-Time-Click-Through Agreements (JITCTAs) \cite{patrick2005}   \\ 
		
		63. Added-noise for obfuscation                                                    & Before sending the user’s data to the system, add fake parking locations so no additional personal information can be inferred by tracking the data collected from the same user.       \\ 
		
		%
		%
		%
		%
		%
		
		70. Active Broadcast of Presence                                                           &                               Before sending car's location and parking data to the server, remind the user of the data being sent and the settings he can change.                             \\ 
		
		%
		%
		%

		\hline
		
	\end{longtable}
	
}

\subsection{Use Case 2: Diabetes Treatment Monitoring}

\begin{figure}[!h]
	\centering
	\includegraphics[scale=.32]{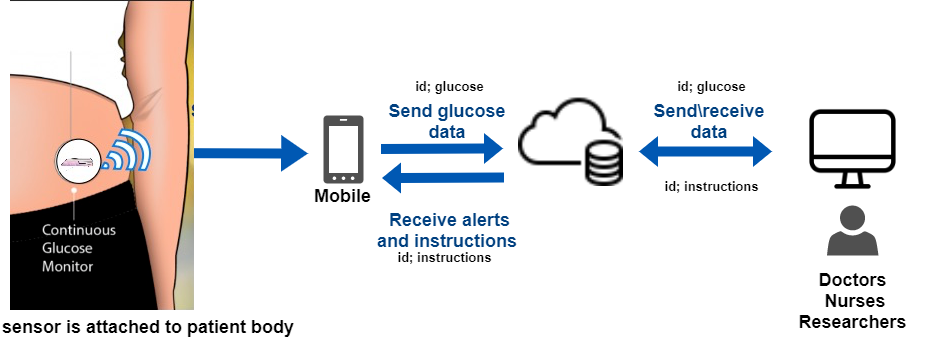}
	\caption{Use case scenario: \textit{Diabetes treatment and monitoring}, IoT application to support diabetes treatment monitoring}
	\label{Fig:CGM_uscase}
\end{figure}

\textit{Diabetes treatment and monitoring} is an IoT application that analyses patient health data to issue alerts and notifications. For simplicity and ease of understanding, this case study is presented from the perspective of a researcher in a healthcare company. The healthcare company has many patients, and one of the cases has diabetes, which requires both treatment and hourly health monitoring. As seen in Figure \ref{Fig:CGM_uscase}, the researcher needs to gather and analyse data from a Continuous Glucose Monitor (CGM) sensor devices worn by her patients. The sensor measures glucose levels by taking readings at consistent intervals across several days.

The researcher uses an application that can detect any triggers for blood glucose levels. This application analyses the gathered data and produces a notification to both the patient and the nurse. In addition to the researcher, the nurse and the doctor have access to patient data for follow-up purposes. If there is a need, the patient could, for example, be instructed to adjust their insulin dosage, do exercise, or change medication. \textcolor{blue}{Health care IoT applications are one of the main focal points in the IoT field, and they also collect very sensitive personal data about their users. In addition, it is a complicated kind of IoT application, which often includes many sensors and requires significant volumes of information. As a result, they are certainly ideal candidates for schemes aiming to preserve privacy, and as such, offer an ideal example for this work. } Table \ref{Tbl:diabetesPatterns} explains how each privacy pattern could help to reduce privacy risk in the \textit{ diabetes treatment and monitoring} IoT application.

{\tiny
	\begin{longtable} {p{4cm} | p{7.3cm}}
		\caption{Application of Privacy Pattern to \textit{diabetes treatment and monitoring} IoT Application}\\
		\hline
		\label{Tbl:diabetesPatterns} 
		Privacy Pattern List &  Example                                                                         \\ 
		\hline
		1. Protection against Tracking                                                       &    Because the system could follow the user behavior, the system should clear the cookies frequently.       \\ 
		\arrayrulecolor{black} 
		2. Location Granularity                                                              & The patient should have the choice to share the level of location details.     \\ 
		
		3. Minimal Information Asymmetry                                                        & The system should only collect needed data, i.e. after analyzing the glucose levels, it only sends the results to the cloud environment not the actual patient information. In  addition,  data  collected,  and  privacy  policies  should  be clearly known to user.  \\ 
		
		4. Informed Secure Passwords                                                            &  The user should be assisted to maintain a strong password to protect themselves. \\ 
		
		5. Awareness Feed.                                                                      & The user should be informed before the application processes the collection of the personal info such as telling the user before sending username and DoB etc. \\ 
		
		6. Encryption with user-managed keys                                         &            This is system involves user password to sign in,secure off-site backup system with client side encryption is needed such as using least Authority.    \\ 
		
		
		8. Use of dummies                                                                      & The dummies are used for personal data, so the system should vague the user’s data by adding fake glucose reading  in the database. \\ 
		
		9. Who’s Listening                                                                      &      User should be able to know who have the access to view their data, i.e. the patient should know that the nurse and research company have accessed the patient readings.   \\

		10. Privacy Policy Display                                                                 &    
		The system should allow users to display the privacy policy at the beginning. \\ 
		
		11. Layered Policy Design                                                                  &   
		The privacy policy should be summarized in a nested way, so patients can understand easily how their data is managed and for which purposes.  \\ 
		
		12. Discouraging Blanket Strategies                                             &                 Users  should have the ability to define a privacy level for content being shared with the research company, doctor or nurse such as sharing his ethnicity group.  \\ 
		
		
		14. Asynchronous notice                                                                     &  Each time patient data is being sent a notification should be given to the user.  \\ 
		
		15. Abridged Terms and Condition                                                           &     
		he privacy policy should be summarized in a concise way, so users can understand how their data is managed and for which purposes.     \\ 
		
		16. Policy Matching Display                                                                  &  The system should retrieve the user’s privacy preferences, via some controller,  and  compare  them  to  its  own  policies,  then,  the  contradictions should be highlighted to the patient.  \\ 
		
		
		18. Outsourcing [with consent]                                                            &            The controller should obtain an additional law consent from users before processing their health data to third parties such as sharing with other research companies.   \\ 
		
		19. Ambient Notice                                                                         &  Each time patient data is being sent a notification should be given to the user.  \\ 
		
		20. Discouraging Privacy Policy                                                           &    In the user interface, a tool tip should be designed, so the user can have a quick glance to the related policy in this input/ service in the interface.  \\ 
		
		21. Privacy Labels                                                                         & In the privacy policy documents, the policies should be explained in a standardised form which will make it easier to the patient to understand. \\ 
		
		22. Data Breach Notification Pattern                                                    & The application should react to data breach quickly by notifying the user with the Breach details. \\ 
		
		23. Pseudonymous Messaging                                                                 &  When exchanging information between the patient and the nurse, the patient’s identity should be pseudonym. \\ 
		
		24. Onion Routing                                                                          &  Transferred data should be encrypted in layers, each edge decrypt a layer. \\ 
		
		
		26. Pseudonymous Identity                                                                 & From researcher perspective, he does not need to know the patient identity, the researcher should only have access to the readings and other impersonal data, such as, age and exercising values.  \\ 
		
		27. Personal Data Store                                                                   &    Patient should be given an access to his personal token and manage his own data.   \\ 
		
		28. Trust Evaluation of Services Slides                                                    &    The system should apply a trust evaluation function that informs users of the trustworthiness and reliability of services        \\ 
		
		29. Aggregation Gateway                                                                   &   The system should apply kind of   encryption such as homomorphic encryption. The encrypted information from patients are transmitted to an independent trusted third party for doing aggregation process (different aggregation result based on the receiver: nurse vs. researcher).

                   \\ 
		
		30. Privacy icons                                                                          &    The privacy polices document should afford standardized visual icon sets beside text which will help park visitors to understand the policies easier.   \\ 
		
		31. Privacy-Aware Network Client                                                         &  Provide a privacy preserving proxy that parses the privacy policies in an easy and understandable format to the patient.  \\ 
		
		32. Sign an Agreement              & The patient should be prompted to sign a straightforward and clear contractual agreement.  \\ 
		
		33. Single Point of Contact                                                              &   The system should provide a Single Point of Contact (SPoC) approach to protect the stored sensitive medical data for the patient \\ 
		
		34. Informed Implicit Consent                                                            &   The patient should be sufficiently informed of all data that are collected about him via the application. \\ 
		
		35. Enable/Disable Function                                                            &  The patient should be able to choose what functions they agree to use via the application.  \\ 
		
		36. Privacy Color Coding                                                                   &    In the privacy policy documents, the policies should be designed with standardized color visual cues to make it easier to the patient to understand the policy. \\ 
		
		37. Appropriate Privacy Icons                                                              &   In the privacy policy documents, the policies should be supplied with the appropriate icons to make understanding the policy easier for the patient. \\ 
		
		38. User Data Confinement Pattern                                                          & The glucose readings abnormal level could be calculated at the patient edge instead of sending raw data to the cloud. \\ 
		
		39. Icons for Privacy Policies                                                             &    In the privacy policy documents, the policies should be supplied with icons to make understanding the policy easier for the patient.\\ 
		
		40. Obtaining Explicit Consent                                                             &  The patient should be prompted an explicit consent via the application that sufficiently explains the consequences of providing their data, the explicit consent, for example, should be asked as “I accept” button. \\ 
		
		41. Privacy Mirrors                                                                      &  Each time the system sends patient’s data, a notification should be shown to the patient. In addition, the user should be given a choice to decide how much and how often privacy details and notifications he would like to see.  \\ 
		
		42. Appropriate Privacy Feedback                                                           &   Each time the system sends patient’s data to the cloud, a notification should be shown to help the patent to understand the processed data about him.  \\ 
		
		
		
		45. Platform for Privacy Preferences                                                   &  Applications are encouraged to use similar privacy preferences platforms, so policy distinctions will be highlighted easily by the users.  \\ 
		
%
		
		48. Privacy Dashboard                                                                  &    Via the application, the patient should be able to view all his collected data easily in summarized design. \\ 
		
		
		50. Obligation Management                                                             &   The system should use an obligation management system, which manipulates data over time, ensuring data deletion and notifications to the user. \\ 
		
		51. Informed Credential Selection                                                     & In the application, the patient should be shown a summary of the choices he has, and the data collected about him before being sent to the cloud. \\ 
		
		
		53. Negotiation of Privacy Policy                                                    & The privacy preserving settings should be applied by default, then, the application can view the options to the patient where he decides to accept them or not based on patient’s preferences. \\ 
		
		54. Reasonable Level of Control                                                     &  In the application interface, allow the patient to choose the granularity of data being sent to the server, for example, he has the choice if he wanted to share what type of exercise he practices. \\ 
		
		
		
		57. Privacy Awareness Panel                                                          & In the application, the patient should be clearly aware that his reports are sent to researchers.  \\ 
		
		58. Lawful Consent                                                                   &  The patient should provide a consent if he wants to share his reports with the researcher or not.  \\ 
		
		59. Privacy Aware Wording                                                                  & In the privacy policy documents, the policies should be explained in an understandable level for the patient.  \\ 
		
		60. Sticky Policies                                                                   & The system should stick to the policies although it shares data with third parties. \\ 
		
		61. Personal Data Table                                                             &  In the application, the patient should be given an access to view the data collected about him in a clear tabular format. For example, glucose readings should be displayed in a table. \\ 
		
		62. Informed Consent                                     &  To make understanding the policies and data processed easy for the patient, apply a Human Computer Interaction concepts, for example, Just-In-Time-Click-Through Agreements (JITCTAs) concept.  \\ 
		
		63. Added-noise  obfuscation                                              &     Add false glucose reading values to the patient records that would be cancelled automatically in a long term. \\ 
		
		64. Increasing   Aggregation Awareness                                  &    The system should inform and apply aggregation as much as possible, for example, the times for the readings could be aggregated in time slots instead of collecting exact time. \\ 
		
		65. Attribute Based Credentials                                                            & The system could only send the attribute value (Normal, abnormal, etc…) of glucose reading percentage. \\ 
		
		66. Trustworthy Privacy Plug-in                                                       & The glucose readings abnormal level could be calculated at the patient edge instead of sending raw data to the cloud. \\ 
		
		
		
		69. Anonymity Set                                                                           &  The system should apply anonymity set mechanism such as, Mix zone, which is a location-aware application that anonymous user identity by limiting the positions where users can be located.  \\ 
		
		
		71. Unusual Activities                                                                &   The system should alert the patient if there is unusual activity, such as, a new mobile login or sudden location change.  \\ 
		
		
		
		\hline
		\arrayrulecolor{black} 
		
		\hline		
	\end{longtable}
}


\section{Discussion}
\label{sec:DiscussionandLessonsLearned}

This section collates and contextualises the results achieved through privacy pattern versus PbD scheme analysis. The outcomes of the raw data analysis are presented in \cite{Alkhariji2020}.

\subsection{Higher Abstraction Mean Easy to Understand}
During the synthesis process, it was realised that, when a given privacy-preserving idea is high-level (Abstract), it is easy to understand (has meaning). For example, privacy principles such as \textit{`Proactive not Reactive; Preventative not Remedial'} are easy to understand compared to a privacy pattern such as \textit{'Minimal Information Asymmetry'}. Furthermore, it is easy to determine how to implement a privacy pattern when compared to a high-level principle. Again, privacy principles such as \textit{`Proactive not Reactive; Preventative not Remedial'} are easy to understand, but require significant thought about how to implement them, especially when a given principle could be interpreted and implemented in many different ways.

\subsection{Patterns can be Explained using High-level PbD Schemes}
We found that high-level PbD schemes could be used to explain how a given privacy pattern would help to reduce privacy risk in a given IoT application design. As discussed above, our approach helps to combine the best of both worlds (i.e., Elements within PbD schemes are easy to understand, and privacy patterns are easy to implement). This means we can use PbD schemes as proxies to explain how a given privacy pattern works.

\subsection{Synthesise Process is Subjective: Crowdsourcing could be a Way-out}
During the PbD knowledge synthesising process, it was noted that some decisions were subjective and could be interpreted in different ways. For example, the question \textit{'Does this privacy pattern follows this [principle, guideline, strategies]'} showed that some answers were subjective and dependent on the method of interpretation of both the pattern and the PbD scheme element in question. \textcolor{blue}{ For instance, the privacy pattern \textit{ 2. Discouraging Blanket Strategies} was categorised against Hoepman’s guidelines \textit{2. Hide} and \textit{ 6. Control} as precisely and partially relevant respectively. It was first asserted that the ability of a use to control who views their data satisfies the \textit{ 6. Control} element, yet it only meets the criteria for \textit{2. Hide} if the service provider does not allow for data sharing between users, since the PbD scheme itself contains abstract vagaries which in themselves are liable to lead to different interpretations. It was concluded that the best approach to addressing this issue is to use crowdsourcing techniques. This means many privacy experts could submit their opinion about a relationship between a given privacy pattern and a given PbD scheme, allowing for a consensus to emerge.}

\subsection{Coverage can tell a story, but not the whole truth}
During PbD knowledge synthesis, several tables were built to demonstrate how different privacy patterns are built by combining the different privacy-preserving ideas presented in alternate PbD schemes, as once in a table format, it is easier to highlight how much coverage each pattern has (e.g., how many PbD ideas each pattern follows). In general, privacy patterns that cover more PbD scheme elements are better. However, caution is advised before saying that coverage alone does not indicate the supremacy of a given privacy pattern. In the next section, several approaches are presented whereby priority can be given to patterns wherein coverage remains the sole concern, though it is admissible that this is not the recommended approach.

\subsection{Simplifying the description of privacy patterns}

During the process of examining the privacy patterns, it was found that some of the patterns were unclear. For instance, the \textit{Minimal Information Asymmetry} pattern is one of the most knotted patterns, where its title says that it is concerned with minimising data collection and discovery. However, after looking at the description of the pattern, it was seen to suggest that the solution arises from two different angles. First, it suggests requesting minimal information from the data subjects so that only as much personal data as is required is collected. Secondly, it reduces the imbalance of policy knowledge by writing clear and concise policies rather than, or in addition to, complex and verbose ones. This type of bagginess occurs in some other patterns as well. However, even the slightest complexities could deter software developers, and thus it is important to write these patterns using a concise, definite vocabulary for ease of universal comprehension.

%
%
%

%

\subsection{Collapse and redundancy among privacy patterns}
	The privacy patterns considered in this paper are collected from different projects (i.e. privacypatterns.eu and privacypatterns.org \cite{privacypatterns.eu,privacypatterns.org}). Each project has its own classifications and categories. During the preparation of this research, it was necessary to merge patterns by removing duplicates across the two projects. The argument is posited that it would be useful to have a unified approach to produce, capture, structure, formulate, and disseminate privacy patterns. Ontologies provide such unification and standardisation capabilities.

\subsection{Well organized resource for privacy patterns}
	
	The privacy patterns considered in this work are collected from different websites (i.e. privacypatterns.eu and privacypatterns.org \cite{privacypatterns.eu,privacypatterns.org}). Each website has its own design and category. In addition, both websites have their own way of organising and categorising the privacy patterns. In the course of this work, it was necessary to identify and merge duplications in the list of identified themes to create a unified list of distinct patterns in a consistently legible format. It is intended that having one resource not only makes combining the patterns easier, but also designs them in a format that makes it easier to retrieve and cluster patterns according to their category.

\subsection{Terminologies used across PbD schemes}
	
The PbD schemes discussed in this paper were developed by different groups from various backgrounds and expertise. This has led to ambiguity in the terminology used to describe elements within them, with terms such as, demonstrate and privacy compliance, accountability and integrity being very close in meaning. This kind of semantic issue can be mitigated by developing an ontology-based knowledge model where similarities can be directly defined from the knowledge base itself.

\begin{figure}[h!]
	\centering
	\includegraphics[scale=0.74, angle =90]{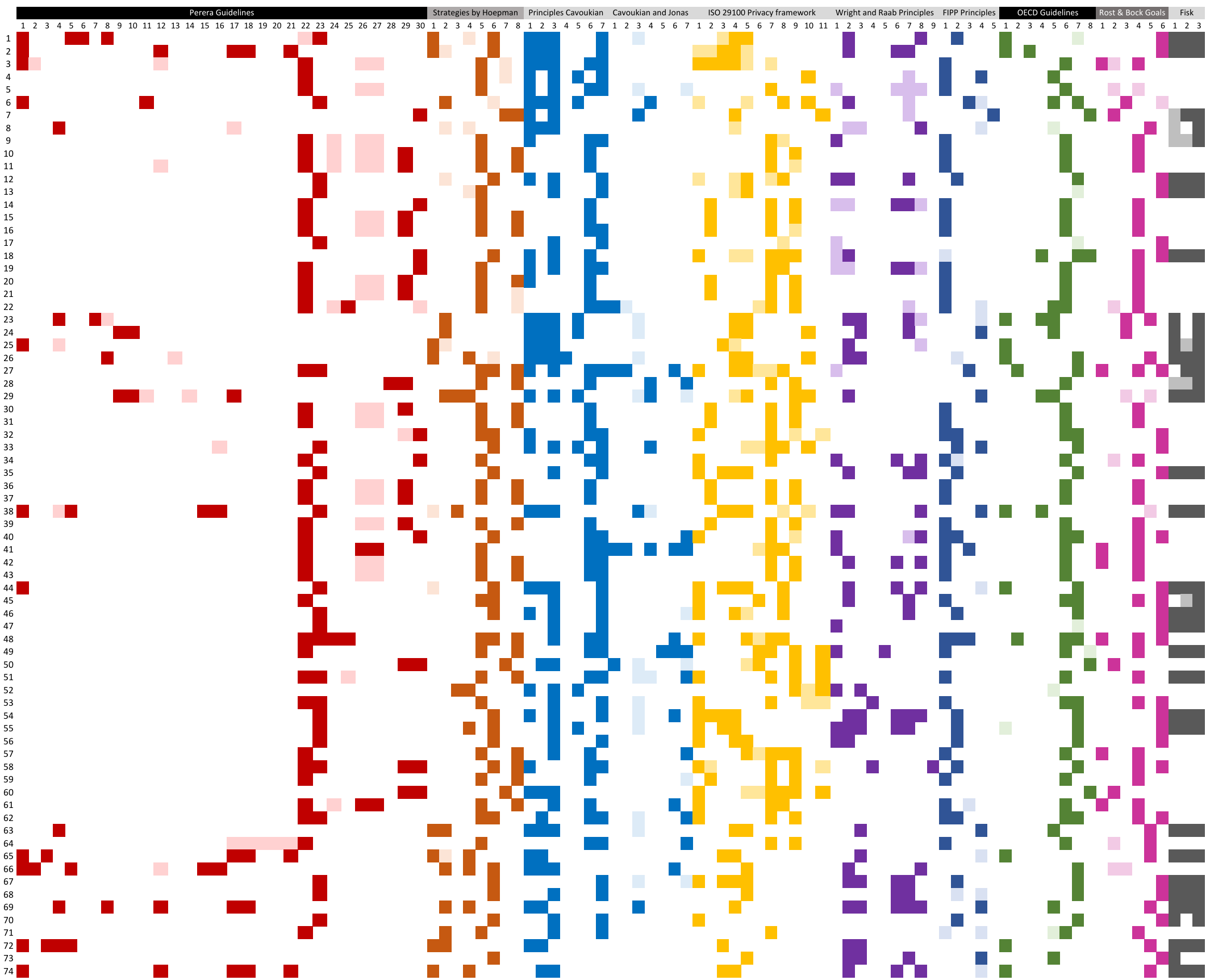}
	\caption{Summarised visualisation of the relationships between privacy patterns and the elements of different PbD schemes. 
	  Heavy and Light colours mean that a particular pattern is fully and partially inspired by a given element within each PbD scheme respectively. The outcomes of the raw data analysis are presented in \cite{Alkhariji2020}.}
	\label{Fig:PattersVsPbDSchmes}	
\end{figure}

\section{Research Challenges and Future Directions}
\label{sec:ResearchChallengesandFutureDirections}

As mentioned earlier, the goal of this paper is to determine requirements for developing an intelligent privacy assistant that can augment the capabilities of software developers at the design stage. This section discusses six identifiably unique challenges that must be addressed in order to develop an intelligent privacy assistant, and the role that a PbD knowledge model would play in this context.

\subsection{Explainable Privacy Assistant}
\label{subsec:ExplainabilityofPrivacyAssistant}
Software developers race against time to deliver their solutions. That places their primary focus on developing functional requirements. If non-functional features such as privacy are considered a necessary part of a successful implementation, then it is imperative that they are provided with appropriate motivators to encourage this incorporation. More recent regulatory efforts, such as GDPR \cite{EUROPEANCOMMISSION2012}, have provided considerable impetus to the conversation, although this investigation confirmed that laws alone do not motivate developers to adopt a PbD approach, and that instead they still require proper guidance, for which they currently have little in the way of support. Either they need to read and gain knowledge about privacy themselves, which conflicts with their core aim of delivering functional requirement as soon as possible, or they need to hire a privacy expert. Depending on the availability and cost, companies may need to rely on a certified privacy professional (iapp.org) or privacy lawyers. However, this is not always possible due to cost constraints. This is especially true for small companies with few developers, who cannot afford to hire privacy lawyers to review their design documents. It is believed that a privacy assistant could help in such situations.

It is important to note that the objective of a privacy assistant is not to replace the job of privacy professionals. Instead, such a privacy assistant could help developers when hiring a privacy professional is not a viable option. Software developers, like any other job role, have their own tacit knowledge involved where over time, they learn to do things in their own ways. As a result, software developers typically do not like to change or be told what to do. Therefore, an ideal privacy assistant should not try to prescribe any privacy-preserving ideas to developers. The ideal task of a privacy assistant is to explain how a given IoT application could be improved by incorporating a set of privacy-preserving ideas. More importantly, developers want to know how a given privacy-preserving idea could help improve their design. To do this, the privacy assistant needs to provide evidence and show how its recommendations help to reduce privacy risk. The synthesis contained herein aims to do just that.

For example, it was discovered that privacy patterns can be explained with the help of other, more abstract PbD schemes (e.g., principles, guidelines, strategies). Even though principles, guidelines, strategies are too abstract and high-level to be useful for implementation, they are relatively easy to understand by a non-privacy expert (e.g., software engineer). For example, earlier in this paper, it was shown that the ’Location Granularity’ privacy pattern follows the minimise strategy. As a result, how ’Location Granularity’ can be used to reduce privacy threats by referring, can be explained through the minimise strategy (See Figure 5). It is believed that it is always important for a privacy assistant to be transparent and not simply prescribe, but also show how it arrived at the recommendation given, by referring to the knowledge it used to make these recommendations. In order to fully realise this ambition, the progress made herein must extend further in the direction of a unified PbD model which can be referenced by future intelligent privacy assistants.

\begin{figure}[h!]
	\centering
	\includegraphics[scale=0.65]{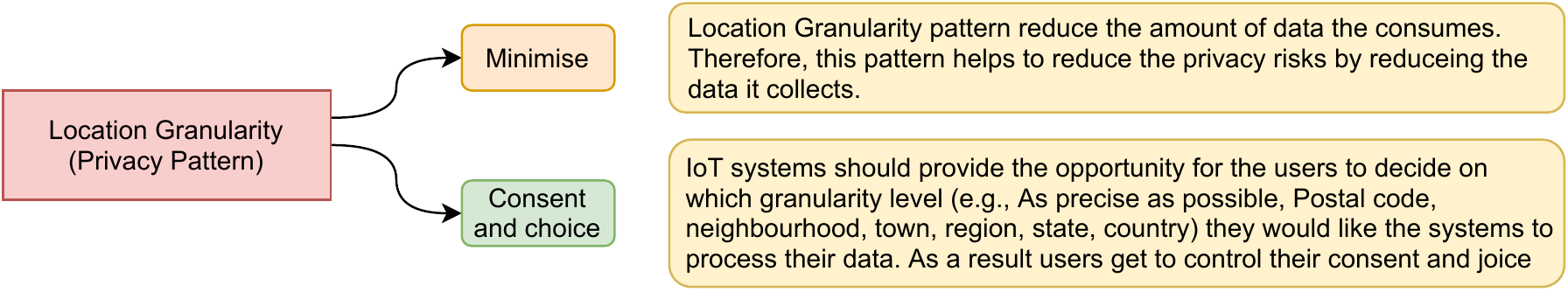}
	\caption{Privacy Patterns could be explained with the help of other high-level PbD Schemes such as principles, guidelines, strategies.}
	\label{Fig:ExplainabilityofPrivacyAssistant}	
\end{figure}

\textcolor{blue}{ More precisely, the IoT scenario we reviewed in \ref{sec:PrivacybyDesignforInternetofThings}—Use Case 2: Diabetes Treatment Monitoring, applies a set of privacy patterns as in table \ref{Tbl:diabetesPatterns}. Those patterns are synthesized to other different PbD schemes as in tables \ref{Tbl:ISOPrivacyPrinciples} and \ref{Tbl:HoepmanPrivacyStrategies}. For instance, The patterns, \textit{10. Privacy Policy Display}, \textit{11. Layered Policy Design} and \textit{16.Policy Matching Display} are a lower level of the ISO principle \textit{7. Openness, transparency and notice} and to Hopeman’s guideline \textit{5. Inform}. Thus, the non-privacy expert software developer will be able to understand what the purpose of those patterns. }

\par 
\textcolor{blue}{The privacy assistant proposed herein will not only help in explaining privacy patterns to software developers, but also to highlight privacy issues that a software developer should consider. This assistant will be used during the design phase. It will be able to take the IoT software design as an input, then, with the help of the built-in knowledge base it will process the software design to find areas of higher risk. Finally, it will show the software developer its privacy concerns and what schemes should be applied to address them.
}

\subsection{Developing a Quarriable Unified PbD Knowledge-base/ontology}
\label{subsec:DevelopinganQuarriablePrivacybyDesignKnowledgebase}

The problem with the current situation is that several different PbD schemes have been developed by different researchers at different time frames with different intentions and mind-sets. There is not a natural connection between them, which is in stark contrast to the privacy patterns built from software design patterning. Although the initial foundations have been laid herein, more work remains to be done to synthesise PbD knowledge and create a unified PbD ontology, and without a unified model, it will remain difficult to make use of all the PbD schemes in an efficient and effective methodology. \textcolor{blue}{ It is necessary to model how each of the elements within each PbD schemes relate to each other. Regular knowledge base representations (i.e. databases) provide less retrieval functionalities and linkabilities between knowledge entities. To build such an expanded knowledge base an intelligent knowledge modelling technique is required, for which it is believed that semantic web technologies will be found appropriate.} \par

An ontology is a formal description of knowledge as a set of concepts within a domain and the relationships that hold between them. Ontology modelling languages such as OWL \cite{P378}, allows modelling rich and complex knowledge about things and the relations between them. One of the main features of ontologies is that, by having the essential relationships between concepts built into them, they enable automated reasoning about data. If we can model PbD knowledge using an ontology modelling language such as OWL, we can use query languages such as SPARQL \cite{P378} to ask related questions, which is ideal to be used for developing a privacy assistant. Such unified PbD ontology will be able to answer questions such as \textit{How Location Granularity pattern helps to reduce privacy risks?}. Due to the graph nature of ontologies, algorithms can move from one concept to the other and discover relationships between different entities and use such information to come up with an explanation autonomously.  \par

\textcolor{blue}{There are forward steps in using ontology as privacy conceptualization tool as well as IoT. For example, Ontology Design Pattern (ODP) (ontologydesignpatterns.org) \cite{Gangemi2009} witch conceptualizes privacy laws and Sensor and Sensor Network ontology (SSN) (w3.org/TR/vocab-ssn/) witch conceptualizes IoT sensors. Adapting such ontologies with AI technology will deliver an intelligent tool that matches the privacy needs of a given IoT software.}

\subsection{Capturing the Context Awareness}
\label{subsec:CapturingtheContextAwareness}

\begin{figure}[h!]
	\centering
	\includegraphics[scale=0.50]{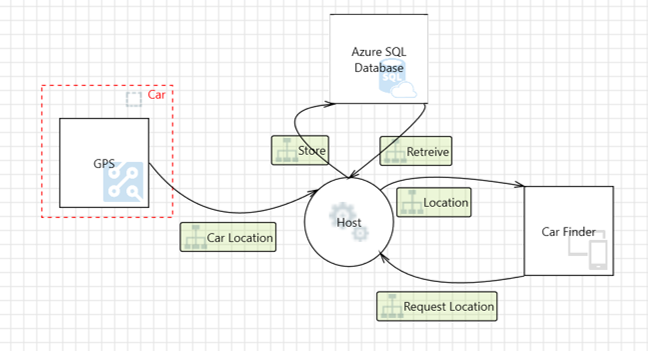}
	\caption{Capturing context information could help to develop a more useful Privacy Assistant that can answer complex queries.}
	\label{Figure:CapturingtheContextAwareness}	
\end{figure}

\textcolor{blue}{Figure \ref{Figure:CapturingtheContextAwareness} presents a part of an IoT thread modelling diagram. It provides information about location tracking IoT application. In this example, the application design refers to GPS sensor. This means that further knowledge can be added about IoT infrastructure, data analytics, and domain knowledge, in order to support more complex and useful queries. A sample query would be \textit{`What are the privacy patterns that should be implemented in a GPS gateway which has only 1GB of memory’}.
}

It is important to note that the novelty of this technique is reliant on the integration of privacy assistance in existing, or new, design tools such as Microsoft’s Threat Modelling Tool \cite{Microsoft2019}.  As a result, software designers can get benefit from this PbD knowledge, and the privacy assistant will develop further on top of that. However, in order to answer more complicated queries like the one posited above, it is not only necessary to model additional information such as memory requirements but also to use rule-based reasoning techniques such as SWRL rule sets \cite{P378}. However, the advantage is, once model, such knowledge can be reused across many different IoT application designs. The challenge is to develop abroad ontology model that goes beyond the PbD Knowledge, that can capture a wide range of aspect mentioned above. We believe that an ontology engineering methodology such as Neon \cite{Figueroa2009} could be ideal for this task. Neon is a scenario-based ontology modelling methodology where you first gather a variety of different questions that the ontology should answer. Then iteratively add more concepts and relationships to enrich the model until all the required questions can be answered while maintaining the consistency.

\subsection{Mediating Lawyer-Developer Communication}
\label{subsec:MediatingLawyerDeveloperCommunication}

It has already been highlighted that firms with larger financial freedoms can simply allocate funds for project inspection by privacy lawyers. However, in practice, this is not an easy process because communication between privacy lawyers and software developers is not typically smooth. Most privacy professionals do not understand software development processes and terminologies and vice versa. It is proposed that PbD knowledge could help to smooth these interactions. The principal problem can be thus considered as the gap between these too stances, which comprises differences in vocabularies, design diagrams and patterns, checklists, perceived priorities, guidance, tools, time frames and methods of working. The ideal response involves extending the above-mentioned ontology model in such a way that it can translate any domain-specific terminologies to those that each party uses. For example, privacy layers use certain vocabulary such as data controller, or data subject, yet from the developers’ point of view, they are dealing with services and users. Therefore, an intermediary knowledge-base can be used to provide efficient and effective communication between the two parties.
%

\subsection{Privacy Patterns Development Methodology}
\label{subsec:PrivacyPatternsDevelopmentMethodology}

Although it was possible to analyse each privacy pattern and examine how they were built using PbD schemes (reverse engineering), privacy patterns are still not being developed using such an approach. Having said that, once a unified PbD knowledge model has been built, a privacy pattern methodology could be developed, which could guide anyone who is interested in developing a new privacy pattern. Current PbD knowledge comprises of significant repetition, yet it is possible to avoid such duplication of work using a unified PbD knowledge base that anyone can query and check whether a similar pattern has been proposed before. Additionally, it could be used to create novel combinations of extant PbD schemes with the intent of identifying new patterns. For example, a strategy such as minimise could inspire the development of new privacy patterns, which are fully inspired by and grounded in previous work.

\subsection{Quantifying the Value of Privacy Patterns}
\label{subsec:QuantifyingtheValueofPrivacyPatterns}
As mentioned throughout this paper, there are many different privacy-preserving ideas proposed by researchers using many different terminologies such as guidelines, principles, strategies, patterns, and goals. The previous discussion showed that motivating software developers to adopt PbD is a challenge. Therefore, if it is proposed to them that they must adopt a large number of privacy-preserving ideas, they are likely to be confused, frustrated and annoyed. A potential outcome of such a situation is that developers will ignore privacy altogether. Therefore it is necessary to find a way to prioritise privacy-preserving ideas. One way to do this is to quantify them. Once quantified, it becomes possible to prioritise and inform developers which privacy-preserving ideas they should implement first.

Quantification can be done in many ways, whereas prioritisation is based on which privacy patterns cover the widest range of PbD elements. For example, during synthesis, the number of PbD elements within each scheme was counted, so that both the raw number and the percentage can be considered. For example, the Location Granularity pattern covers fours strategies out of eight within Hoepman’s \cite{Hoepmanraey} framework of strategies. Additionally, these numbers can be summed across all ten PbD schemes, and crowdsourcing techniques used to uncover how two elements from two different PbD schemes could be related to each other. It is expected that there could be diverse opinions, but that this method should allow for the consensus to emerge.  

Another type of quantification would be to look at the time needed to implement. However, this would be challenging, as the time to implement varies heavily with both the programming environment and the complexity of a given design. However, such time to implement information could be very useful, because software developers could undertake better planning when they know the approximate time requirements for a project. Another aspect is to prioritise based on the importance. Thus the principal aim is to determine the preservation method with the highest yield. One way to quantify this is to look at past privacy violations and the damage they have caused. Then, choose the privacy pattern that could have avoided this scenario, and score each idea based on the added value provided through adoption of the given concept. As software developers are pitted against both time and cost, any decision on what to implement is effectively a multi-criteria optimisation problem and thus will result in compromise. Important criteria to consider would be: (i) best privacy protection coverage, (ii) ease/time/cost of development, and (iii) access to the knowledge that is required to implement a given privacy pattern.

\subsection{Visualizing Privacy}
\label{subsec:VisualizingPrivacy}
\textcolor{blue}{ Visualising privacy could be interpreted in a few different ways. One way is to visualise the relationships between different PbD schemes. In this paper, only the relationships between patterns and other PbD schemes was explored. However, it is necessary to explore the relationships between elements within PbD schemes as well. Such visualisation will create graphics which can be used to study how PbD elements are connected to each other. Another way to explain and visualise privacy patterns is through use cases. For example, in an interactive platform that allows a software developer to learn how privacy preserving ideas are applied in a given IoT application design. The objective here is to help software developers to easily recall what a particular privacy preserving idea achieves.}

\section{Conclusion}
\label{sec:Conclusion}

\textcolor{blue}{ This paper reviews, analyses and synthesises different types of Privacy by Design schemes. Their interrelations are explored through preconceived expectations with the aim of developing a unified PbD knowledge model. Privacy patterns are difficult to understand, but easy to implement, whereas, guidelines, strategies, principles, goals, and rights are the opposite. It was realised that these high-level PbD schemes could be used to describe and explain privacy patterns. More specifically, the findings show that most privacy patterns are composed (or inspired) by more than one privacy by design element. As a result, PbD schemes and their elements can be used to explain how a given privacy pattern is not only relevant to a given IoT application design but can also be used to explain to software developers how they protect privacy. With the help of IoT use-cases, it was possible to extract several knowledge engineering requirements. Finally, three use case scenarios were used to demonstrate how privacy patterns would be useful in designing privacy-aware IoT applications. It was also found that privacy patterns can significantly benefit healthcare applications compared to other IoT application domains. Thus, this work lays the foundation towards developing a privacy assistant by synthesising existing PbD knowledge and defining future requirements.
}

\begin{acks}
Charith Perera's work is partially supported by PETRAS2: National Centre of Excellence for IoT Systems Cyber Security (EP/S035362/1) and Quarriable Smart City Data Markets (EP/T517203/1). Omer Rana's work is supported by PACE: Privacy-aware Cloud Ecosystems (EP/R033439/1). Mansour Naser Alraja is supported by The Research Council (TRC), Sultanate of Oman ( Block Fund-Research Grant).
\end{acks}

\bibliographystyle{ACM-Reference-Format}
\bibliography{library2}

\newpage

\appendix

\section{APPENDIX}

\subsection{Use Case 3: Leisure Park Quality Monitoring}

\begin{figure}[H]
	\centering
	\includegraphics[scale=0.5]{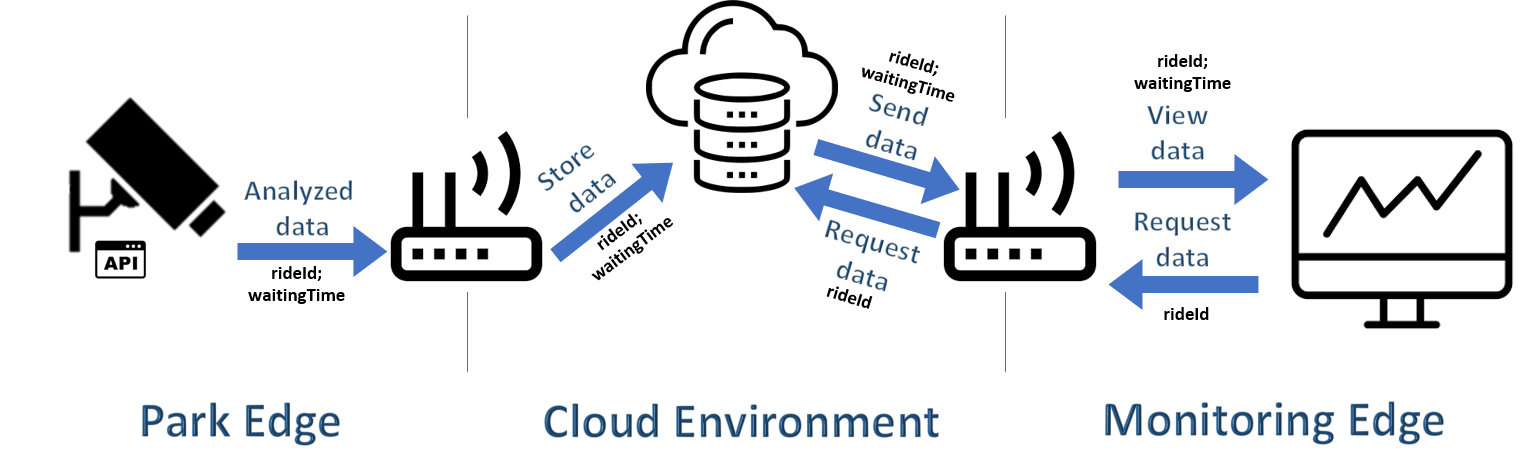}
	\caption{Use case scenario: \textit{TrueLeisure}: Performance Monitoring}
	\label{Fig:TrueLesuireScenario}
\end{figure}

\textit{TrueLeisure} is an IoT application that evaluates the performance of a Leisure park. Waiting time is one of the key service quality attributes and is a key contributory factor to customer satisfaction. Local quality assessment teams continuously measure the crowd waiting time of each ride and attraction within their own amusement park. True Lesuire is a park theme that facilitates many rides for visitors with waiting queue. This system will help to get expected waiting time for each ride. A snapshot for the queue will be taken from a camera device, the photos will be analysed by applying a specific API. Then, the analysed data will be sent from the Park edge to the cloud environment via the gate way. The supervising team are able to monitor the status for each ride from the monitoring edge. They can request and retrieve the data from the cloud environment via the gateway. Figure \ref{Fig:TrueLesuireScenario} illustrates the use case. In Table \ref{Tbl:TrueLeisurePatterns}, we explain how each privacy pattern could help to reduce privacy risk in the \textit{TrueLeisure} IoT application.

Team member get the software in mobile devices that analyses expected waiting time. Only Park name, attraction picture and waiting time are sent to the cloud. Jane gets the Results in the monitoring screen. All photos taken by the mobile application are instantly deleted. Such a common distributive system will help us to understand the needs of such IoT applications.


{\tiny
	\begin{longtable} {p{4cm}  | p{7.3cm}}
		\caption{Application of Privacy Pattern to \textit{TrueLeisure} IoT Application}\\
		\hline
		\label{Tbl:TrueLeisurePatterns} 
		Privacy Pattern List  & Example                                                                         \\ 
		\hline
		\arrayrulecolor{black} 
		
		3. Minimal Information Asymmetry                                                        & The system should only collect needed data, i.e. after analyzing the photos, it only sends the results to the cloud environment not the actual pictures.       \\ 
		
		
		
		
		
		
		
		10. Privacy Policy Display                                                                 &    Before visitors enter the theme park, they should be informed about the privacy policies. \\ 
		
		11. Layered Policy Design                                                                  &   Visitors will not read long documents, so the policies should be given in summaries, highlighting the non-evident ones for them.  \\ 
		
		
		
		14. Asynchronous notice                                                                     &  Cameras in the park should blink a light as a notification each time it takes a snapshot.  \\ 
		
		15. Abridged Terms and Condition                                                           &      The privacy policies document should be written to the park visitors perspective focusing on the most important matters. Such as, highlighting that the photos taken are not stored and only the analyzing results are sent to the cloud environment.     \\ 
		
		
		
		
		19. Ambient Notice                                                                         &  Cameras in the park should blink a light as a notification each time it takes a snapshot. \\ 
		
		
		21. Privacy Labels                                                                         & In the privacy policy document, the policies should be in standardized privacy labels to help users understand them with less effort. \\ 
		
		
		
		24. Onion Routing                                                                          &  Encrypt the data in layers, then, when transferring data between edges, each edge can remove one layer of encryption.  \\ 
		
		
		
		
		28. Trust Evaluation of Services Slides                                                    &    The system should supply a trust certificate, provided by EuroPrise for example, that will make park visitors comfort satisfied.        \\ 
		
		
		30. Privacy icons                                                                          &    The privacy polices document should afford standardized visual icon sets beside text which will help park visitors to understand the policies easier.   \\ 
		
		
		32. Sign an Agreement              & Park visitors should be provided with a clear, straightforward contractual agreement document which the system is obligated to.   \\ 
		
		
		
		
		36. Privacy Color Coding                                                                   &    In the privacy policies document, the policies should be designed with standardized color visual cues to help park visitors to understand them easily.  \\ 
		
		37. Appropriate Privacy Icons                                                              &   In the privacy policies document, support the policies with appropriate visual icons that will help park visitors to understand them easily.  \\ 
		
		38. User Data Confinement Pattern                                                          & The system analysis the data from the photos in the park edge, it does not send the photos to the cloud environment.  \\ 
		
		39. Icons for Privacy Policies                                                             &    In the privacy policies document, support the policies with visual icons that will help park visitors to understand them easily. \\ 
		
		40. Obtaining Explicit Consent                                                             &   The theme park is responsible to provide enough explanation for visitors and prompt them an explicit consent.  \\ 
		
		
		42. Appropriate Privacy Feedback                                                           &   Near each camera in the theme park provide a visible note that reminds park visitors about the camera in the zone.  \\

		59. Privacy Aware Wording                                                                  & In privacy policies document, used vocabulary should be easy to understand by any user, considering applying other patterns as well, such as, layered policy design.  \\ 
		
		%
		%
		%
		%
		
		65. Attribute Based Credentials                                                            & The monitoring system does apply this pattern, it does not store raw personal data, it only sends photos’ analysed results to the cloud environment. \\ 
		
		%
		%
		
		69. Anonymity Set                                                                           &  The system does not store park visitors’ identities nor photos, it only concerns about the waiting time in the queue.  \\ 
		
		
		
		
		%
		\hline
		\arrayrulecolor{black}

	\end{longtable}
}

\subsection{Relationships Between Privacy Patterns and PbD Schemes}

\begin{description}
	\item[\textbf{PbD Guidelines by Perera et al. \cite{Perera2017}}] 
	These guidelines were the only scheme that was specially designed to assess IoT software and IoT middleware platforms. Therefore, the guidelines provide a satisfying level of details about how to improve the data subject’s privacy level in IoT applications. Most of the guidelines were able to be matched with privacy patterns. The guideline \textit{11. Encrypted data storage} is an example, where it can be directly mapped to privacy pattern \textit{6. Encryption with user-managed keys}. This pattern is also applying the strategy by Hoepman \textit{2. Hide}.

	Perera’s guidelines are derived from Hoepman’s strategies. However, it does not mean that the relationships between strategies and guidelines are valid when they are mapped to privacy patterns. For example, pattern \textit{3. Minimal Information Asymmetry} does not follow \textit{1. Minimise strategy} but it does follow \textit{1. Minimise data acquisition} guideline. However, pattern \textit{3. Minimal Information Asymmetry}, which is categorized under \textit{Inform} category, implicitly minimize data acquisition, while \textit{1. Minimise strategy} is directly focusing on tightening data collection.

	Perera’s guidelines are fairly specific. Each guideline specifies how to apply within an IoT application scenario. However, as mentioned in Section \ref{sec:2}, guidelines are not necessarily easy to be implemented by software developers. As shown in figure \ref{Fig:PattersVsPbDSchmes}, we notice a lack of privacy patterns that make use of these guidelines. Thus, we suggest that an additional set of privacy patterns could be derived from Perera’ guidelines.
	
\end{description}

\begin{description}
	\item[\textbf{PbD Strategies by Hoepman et al. \cite{Hoepmanraey}}] 
	Hoepman’s eight strategies are either data-oriented strategies or process-oriented strategies. Most of the privacy patterns follow the ideas presented in Hoepman’s privacy strategies. A significant number of privacy patterns follow the ideas presented in \textit{5. Inform} strategy Further, the process-oriented strategies, \textit{6.Control} and \textit{8. Demonstrate} have also inspired a significant number of patterns. The reason is that most of the privacy patterns are concerned about informing data subject, giving him control and explaining the policies for him.

	The privacy pattern \textit{6. Encryption with user-managed keys} is categorized under \textit{Control} category. It does not say that the data subject should be able to edit or delete his data. Instead, it says that the data subject should have the ability to keep data confidential. Hoepman’s idea of control is that the data subject should be able to consent and update his own data. Thus, we can conclude that \textit{6. Encryption with user-managed keys} partially follows the  \textit{Control} strategy with a (\tikzcircle[orange, fill=orange]{1.5pt}) circle.

	The privacy pattern \textit{44. Decoupling [content] and Location Information Visibility} says that data subject should be able to decide if he is willing to disclose location data within the content he is sharing, that should be his own choice. Thus, it follows the strategy \textit{6. Control}. Moreover, it also leads to collect less data. Thus, it partially follows the concept of \textit{1. Minimize strategy}.

\end{description}

\begin{description}
	\item[\textbf{PbD Principles by Cavoukian et al. \cite{Cavoukian}}] 
	
	Cavoukian’s seven principles are considered as the foundational principles that highlight respect for user privacy.  These foundational principles seem to have inspired a significant number of privacy patterns, as shown in figure \ref{Fig:PattersVsPbDSchmes}. However, some principles are hardly followed by privacy patterns. The principle \textit{4. Full Functionality-Positive-Sum, not Zero-Sum} was hardly followed by any privacy patterns. This principle is raising the issue of getting the benefit of the services provided for both the controller and the data subject by pursuing innovative solutions and options.

	The privacy pattern \textit{26. Pseudonymous Identity} states that data should not be linked to an identity where possible, which presents a win-win scenario, where the service is still running, and data subject’s privacy is protected. However, it is worth noting that there are not enough patterns that adopt this way of thinking.

\end{description}

\begin{description}
	\item[\textbf{PbD Principles by Cavoukian and Jonas \cite{Cavoukian2012}}] 
	
	Cavoukian and Jonas's seven principles are a deeper expression of the first Cavoukian's 7 principles. These principles accommodate the sense of responsibility for software developers. Data subjects are not involved in this particular scheme. However, it provides too specific rules to be followed by patterns. For instance, privacy pattern \textit{69. Anonymity Set} states to aggregate a set of data so the individual's privacy would be preserved.  This pattern seems to be inspired by the principle \textit{3. Analytics on Anonymised Data}.
	
	Despite the fact that these principles provide valuable privacy-preserving ideas, base on Figure \ref{Fig:PattersVsPbDSchmes}, privacy patterns seems to have ignored them. We argue that more privacy patterns should be developed inspired by these principles
	
\end{description}

\begin{description}
	\item[\textbf{PbD Principles by ISO 29100 Privacy framework \cite{ISO2015}}] 
	The International Organization for Standardization (ISO), is one of the well known PbD schemes. The 11 principles are extensively covered across the privacy patterns, as shown in Figure \ref{Fig:PattersVsPbDSchmes}. For instance, privacy pattern \textit{38. User Data Confinement Pattern} follows the principles, \textit{3. Collection limitation}, \textit{4. Data minimization} and \textit{5. Use, retention and disclosure limitation}. Further, it also partially follows two principals, \textit{8. Individual participation and access} and  \textit{10. Information security}. 
	
	It is also noticed that, because ISO principles are focusing on law aspects,  terms used to describe privacy concepts could create confusion, especially among developers. However, ISO principles could be used to identify the link between privacy laws and privacy patterns.

\end{description}

\begin{description}
	\item[\textbf{PbD Principles by Wright and Raab  \cite{Wright2014}}] 
	
	Wright and Raab nine principles are formulated in addition to the ISO 11 principles. They find ISO principles focus more on data protection, rather than other types of privacy, i.e. personal dignity and moral practices. As shown in Figure \ref{Fig:PattersVsPbDSchmes}, privacy patterns do not seem to be inspired by some of the principles at all. For example, the principles \textit{4. Right to autonomy}, \textit{5. Right to individuality} and \textit{ 9. People should not have to pay} are rarely followed by privacy patterns matching.

	Wright and Raab’s principles are highlighting a major gap regarding data subject’s rights, not informational protection. There is an opportunity to develop privacy patterns that operationalise some of the idea presented by principles such as 4, 5, and 9.
	
\end{description}

\begin{description}
	\item[\textbf{PbD Principles by Fair Information Practice Principles (FIPPs)  \cite{Cate2006}}] 
	
	Fair Information Practice Principles (FIPPS) were introduced in 1970s. They were broadly accepted and used as a reference law in the United States, Europe, and elsewhere. As shown in Figure \ref{Fig:PattersVsPbDSchmes}, privacy patterns are well inspired by FIPP principles. Since privacy patterns are mostly focused on informing the data subject about the collection/ progression and control on their data, the principles \textit{1.Notice / Awareness} and \textit{2.Choice / Consent} are well followed by the privacy patterns. The privacy pattern \textit{40. Obtaining Explicit Consent}, which encourages not only informing data subject about what and how his data is procced and used but also prompting him explicit permission to do so, is an example of applying both of principles.

	This scheme shows an example of confusion in using terminologies. It might seem at first sight that the privacy pattern \textit{46. Selective Access control} is a pattern that follows the principle  \textit{3. Access / Participation}. However, \textit{46. Selective Access control} pattern states that the data subject should be able to choose 'who' could have access to his own data, while the \textit{3. Access / Participation} principle states that the data subject should be able to access and evaluate his data in the controller's servers. Therefore, developing a semantic model is vital to distinguish these differences.

\end{description}

\begin{description}
	\item[\textbf{PbD Guidelines by (OECD) Oleary 1995 \cite{Oleary1995}}]

	The OECD guidelines are introduced and followed by 30 countries \cite{Wright2011}. They were primarily developed for protecting the privacy of personal data. As shown in Figure \ref{Fig:PattersVsPbDSchmes}, the privacy patterns mostly follow the guidelines \textit{5. Security safeguards}, \textit{6. Openness} and \ textit{7. Individual participation}. Privacy patterns are less likely to follow the other guidelines. The guidelines mentioned above mostly focus on informing the users and giving them control. For example, privacy pattern \textit{48. Privacy Dashboard}, indicates that the data subjects should have a user interface where they could review and modify their collected data. This pattern is applying both guidelines, \textit{6. Openness} and \ textit{7. Individual participation}. On the other hand, the principle \textit{3. Purpose specification} are not well supported by privacy patterns, although it specifies an important aspect in the privacy by design concept which highlight an important area to develop new privacy patterns.
	
\end{description}

\begin{description}
	\item[\textbf{PbD Goals by Rost and Boc \cite{Rost2011}}] 
	
	Rost and Bock have formulated six privacy by design protection goals. The data protection goals have incorporated privacy by design principles and requirements to meet the specific demands of data protection. The first three goals \textit{1. availability}, \textit{2. integrity}, and \textit{3. confidentiality} are formulated for the safe and secure maintenance of operation and infrastructure of an organization. The other three goals, \textit{4. transparency}, \textit{5. unlikability} and \textit{6. ability to intervene} focus on the individual's perspective. 
	
	Since most of the privacy patterns focused on privacy, not security, they seem to follow the goals \textit{4. Transparency} and \textit{6. ability to intervene}, as shown in Figure \ref{Fig:PattersVsPbDSchmes}. Some of the privacy patterns (i.e. the ones categorized under minimizing) seem to follow the unlikability goal. For example, patterns \textit{23. Pseudonymous Messaging} and \textit{26. Pseudonymous Identity}, states that the massages and data should not be linked to the identity of the data subject.  Nevertheless, the goals \textit{2. Integrity} and \textit{3. Confidentiality} are overlooked concepts in the view of privacy patterns that need to be enhanced.

\end{description}

\begin{description}
	\item[\textbf{PbD Principle by Fisk et al.  \cite{Fisk2015}}] 
	
	Fisk et al. have proposed three principles that focus on privacy protection from the controller’s perspective. The principles increase preserving the data subject’s privacy when sharing their information across multiple organizations. As shown in Figure \ref{Fig:PattersVsPbDSchmes}, the privacy patterns that apply these principles are the ones showing minimizing and sharing data with third parties. For example, privacy pattern \textit{21. Privacy Labels} is focused on informing data subject about the privacy policies followed by the controller. These three principles are not supporting any of the informing patterns. The privacy pattern \textit{38. User Data Confinement Pattern} seems to be inspired by all the three principles, where it states not send data subject’s information to the server, except the essential. The three principles seem to be dependable, i.e. if a pattern is following one principle, it likely follows the other two principles as well. The reason is that they are very closely aligned with each other.
	
\end{description}

%
%
%
%
%
%
%

\end{document}